\documentclass[a4paper,11pt]{article}
\makeatletter
\gdef\@fpheader{}

\newif\ifnotoc\notocfalse
\newif\ifemailadd\emailaddfalse
\newif\iftoccontinuous\toccontinuousfalse
\newif\ifnatbibsort\natbibsorttrue

\relax

\RequirePackage{amsmath}
\RequirePackage{amssymb}
\RequirePackage{epsfig}
\RequirePackage{graphicx}
\ifnatbibsort\RequirePackage[numbers,sort&compress]{natbib}\else\RequirePackage[numbers,compress]{natbib}\fi
\RequirePackage[colorlinks=true
  ,urlcolor=blue
  ,anchorcolor=blue
  ,citecolor=blue
  ,filecolor=blue
  ,linkcolor=blue
  ,menucolor=blue
  ,pagecolor=blue
  ,linktocpage=true
  ,pdfproducer=medialab
  ,pdfa=true
]{hyperref}

\def\@subheader{\@empty}
\def\@keywords{\@empty}
\def\@abstract{\@empty}
\def\@xtum{\@empty}
\def\@dedicated{\@empty}
\def\@arxivnumber{\@empty}
\def\@collaboration{\@empty}
\def\@collaborationImg{\@empty}
\def\@proceeding{\@empty}
\def\@preprint{\@empty}

\newcommand{\subheader}[1]{\gdef\@subheader{#1}}
\newcommand{\keywords}[1]{\if!\@keywords!\gdef\@keywords{#1}\else%
\PackageWarningNoLine{\jname}{Keywords already defined.\MessageBreak Ignoring last definition.}\fi}
\renewcommand{\abstract}[1]{\gdef\@abstract{#1}}
\newcommand{\dedicated}[1]{\gdef\@dedicated{#1}}
\newcommand{\arxivnumber}[1]{\gdef\@arxivnumber{#1}}
\newcommand{\proceeding}[1]{\gdef\@proceeding{#1}}
\newcommand{\xtumfont}[1]{\textsc{#1}}
\newcommand{\correctionref}[3]{\gdef\@xtum{\xtumfont{#1} \href{#2}{#3}}}

\newcommand\preprint[1]{\gdef\@preprint{\hfill #1}}

\newcommand\note[2][]{%
\if!#1!%
\stepcounter{footnote}\footnotetext{#2}%
\else%
{\renewcommand\thefootnote{#1}%
\footnotetext{#2}}%
\fi}

\newtoks\auth@toks
\renewcommand{\author}[2][]{%
  \if!#1!%
    \auth@toks=\expandafter{\the\auth@toks#2\ }%
  \else
    \auth@toks=\expandafter{\the\auth@toks#2$^{#1}$\ }%
  \fi
}

\newtoks\affil@toks\newif\ifaffil\affilfalse
\newcommand{\affiliation}[2][]{%
\affiltrue
  \if!#1!%
    \affil@toks=\expandafter{\the\affil@toks{\item[]#2}}%
  \else
    \affil@toks=\expandafter{\the\affil@toks{\item[$^{#1}$]#2}}%
  \fi
}

\newtoks\email@toks\newcounter{email@counter}%
\newcommand{\emailAdd}[1]{%
\emailaddtrue%
\ifnum\theemail@counter>0\email@toks=\expandafter{\the\email@toks, \@email{#1}}%
\else\email@toks=\expandafter{\the\email@toks\@email{#1}}%
\fi\stepcounter{email@counter}}
\newcommand{\@email}[1]{\href{mailto:#1}{\tt #1}}

\newcommand*\collaboration[1]{\gdef\@collaboration{#1}}
\newcommand*\collaborationImg[2][]{\gdef\@collaborationImg{#2}}

\newcommand\afterLogoSpace{\smallskip}
\newcommand\afterSubheaderSpace{\vskip3pt plus 2pt minus 1pt}
\newcommand\afterProceedingsSpace{\vskip21pt plus0.4fil minus15pt}
\newcommand\afterTitleSpace{\vskip23pt plus0.06fil minus13pt}
\newcommand\afterRuleSpace{\vskip23pt plus0.06fil minus13pt}
\newcommand\afterCollaborationSpace{\vskip3pt plus 2pt minus 1pt}
\newcommand\afterCollaborationImgSpace{\vskip3pt plus 2pt minus 1pt}
\newcommand\afterAuthorSpace{\vskip5pt plus4pt minus4pt}
\newcommand\afterAffiliationSpace{\vskip3pt plus3pt}
\newcommand\afterEmailSpace{\vskip16pt plus9pt minus10pt\filbreak}
\newcommand\afterXtumSpace{\par\bigskip}
\newcommand\afterAbstractSpace{\vskip16pt plus9pt minus13pt}
\newcommand\afterKeywordsSpace{\vskip16pt plus9pt minus13pt}
\newcommand\afterArxivSpace{\vskip3pt plus0.01fil minus10pt}
\newcommand\afterDedicatedSpace{\vskip0pt plus0.01fil}
\newcommand\afterTocSpace{\bigskip\medskip}
\newcommand\afterTocRuleSpace{\bigskip\bigskip}
\newlength{\affiliationsSep}
\newcommand\beforetochook{\pagestyle{myplain}\pagenumbering{roman}}

\DeclareFixedFont\trfont{OT1}{phv}{b}{sc}{11}

\renewcommand\maketitle{
\pagestyle{empty}
\thispagestyle{titlepage}
\setcounter{page}{0}
\noindent{\small\scshape\@fpheader}\@preprint\par
\afterLogoSpace
\if!\@subheader!\else\noindent{\trfont{\@subheader}}\fi
\afterSubheaderSpace
\if!\@proceeding!\else\noindent{\sc\@proceeding}\fi
\afterProceedingsSpace
{\LARGE\flushleft\sffamily\bfseries\@title\par}
\afterTitleSpace
\hrule height 1.5\p@%
\afterRuleSpace
\if!\@collaboration!\else
{\Large\bfseries\sffamily\raggedright\@collaboration}\par
\afterCollaborationSpace
\fi
\if!\@collaborationImg!\else
{\normalsize\bfseries\sffamily\raggedright\@collaborationImg}\par
\afterCollaborationImgSpace
\fi
{\bfseries\raggedright\sffamily\the\auth@toks\par}
\afterAuthorSpace
\ifaffil\begin{list}{}{%
\setlength{\leftmargin}{0.28cm}%
\setlength{\labelsep}{0pt}%
\setlength{\itemsep}{\affiliationsSep}%
\setlength{\topsep}{-\parskip}}
\itshape\small%
\the\affil@toks
\end{list}\fi
\afterAffiliationSpace
\ifemailadd %% if emailadd is true
\noindent\hspace{0.28cm}\begin{minipage}[l]{.9\textwidth}
\begin{flushleft}
\textit{E-mail:} \the\email@toks
\end{flushleft}
\end{minipage}
\else %% if emailaddfalse do nothing
\PackageWarningNoLine{\jname}{E-mails are missing.\MessageBreak Plese use \protect\emailAdd\space macro to provide e-mails.}
\fi
\afterEmailSpace
\if!\@xtum!\else\noindent{\@xtum}\afterXtumSpace\fi
\if!\@abstract!\else\noindent{\renewcommand\baselinestretch{.9}\textsc{Abstract:}}\ \@abstract\afterAbstractSpace\fi
\if!\@keywords!\else\noindent{\textsc{Keywords:}} \@keywords\afterKeywordsSpace\fi
\if!\@arxivnumber!\else\noindent{\textsc{ArXiv ePrint:}} \href{https://arxiv.org/abs/\@arxivnumber}{\@arxivnumber}\afterArxivSpace\fi
\if!\@dedicated!\else\vbox{\small\it\raggedleft\@dedicated}\afterDedicatedSpace\fi
\ifnotoc\else
\iftoccontinuous\else\newpage\fi
\beforetochook\hrule
\tableofcontents
\afterTocSpace
\hrule
\afterTocRuleSpace
\fi
\setcounter{footnote}{0}
\pagestyle{myplain}\pagenumbering{arabic}
} % close the \renewcommand\maketitle{

\renewcommand{\baselinestretch}{1.1}\normalsize
\divide\@tempdima\baselineskip
\@tempcnta=\@tempdima
\skip\@mpfootins = \skip\footins
\renewcommand{\@dotsep}{10000}

\newcommand\ps@myplain{
\pagenumbering{arabic}
\renewcommand\@oddfoot{\hfill-- \thepage\ --\hfill}
\renewcommand\@oddhead{}}
\let\ps@plain=\ps@myplain

\newcommand\ps@titlepage{\renewcommand\@oddfoot{}\renewcommand\@oddhead{}}

\numberwithin{equation}{section}

\renewcommand\section{\@startsection{section}{1}{\z@}%
                                   {-3.5ex \@plus -1.3ex \@minus -.7ex}%
                                   {2.3ex \@plus.4ex \@minus .4ex}%
                                   {\normalfont\large\bfseries}}
\renewcommand\subsection{\@startsection{subsection}{2}{\z@}%
                                   {-2.3ex\@plus -1ex \@minus -.5ex}%
                                   {1.2ex \@plus .3ex \@minus .3ex}%
                                   {\normalfont\normalsize\bfseries}}
\renewcommand\subsubsection{\@startsection{subsubsection}{3}{\z@}%
                                   {-2.3ex\@plus -1ex \@minus -.5ex}%
                                   {1ex \@plus .2ex \@minus .2ex}%
                                   {\normalfont\normalsize\bfseries}}
\renewcommand\paragraph{\@startsection{paragraph}{4}{\z@}%
                                   {1.75ex \@plus1ex \@minus.2ex}%
                                   {-1em}%
                                   {\normalfont\normalsize\bfseries}}
\renewcommand\subparagraph{\@startsection{subparagraph}{5}{\z@}%
                                   {1.75ex \@plus1ex \@minus .2ex}%
                                   {-1em}%
                                   {\normalfont\normalsize\itshape}}

\def\fnum@figure{\textbf{\figurename\nobreakspace\thefigure}}
\def\fnum@table{\textbf{\tablename\nobreakspace\thetable}}

\long\def\@makecaption#1#2{%
  \vskip\abovecaptionskip
  \sbox\@tempboxa{\small #1. #2}%
  \ifdim \wd\@tempboxa >\hsize
    \small #1. #2\par
  \else
    \global \@minipagefalse
    \hb@xt@\hsize{\hfil\box\@tempboxa\hfil}%
  \fi
  \vskip\belowcaptionskip}

\renewenvironment{thebibliography}[1]{%
\begin{oldthebibliography}{#1}%
\small%
\raggedright%
\setlength{\itemsep}{5pt plus 0.2ex minus 0.05ex}%
}%
{%
\end{oldthebibliography}%
}
\makeatother

\title{\boldmath   Entanglement Structure of Nonlocal Field Theories }

\author[a,b]{Reza Pirmoradian,}
\author[c]{M. Hossein Bek-Khoshnevis,}
\author[c]{Sadaf Ebadi,}
\author[a,c,d]{M. Reza Tanhayi}

\affiliation[a]{School of Quantum Physics and Matter
	Institute for Research in Fundamental Sciences (IPM),
	P.O. Box 19395-5531, Tehran, Iran}
\affiliation[b]{Ershad Damavand, Institute of Higher Education (EDI)\\
P.O. Box 14168-34311, Tehran, Iran}
\affiliation[c]{Department of Physics, CT.C, Islamic Azad University, Tehran, Iran\\ P.O. Box 14676-86831,
	Tehran, Iran}
\affiliation[d]{Institute of Biosocial and Quantum Science and Technologies, CT.C, Islamic Azad University, Tehran, Iran}
\emailAdd{rezapirmoradian@ipm.ir}
\emailAdd{mh.bkhoshnevis@gmail.com}
\emailAdd{sadafebadi08@gmail.com}
\emailAdd{mtanhayi@ipm.ir}

\abstract{
Nonlocal interactions are known to generate volume-law entanglement entropy. However, their deeper impact on the fine structure of quantum correlations remains a key open question. In this work, we explore a bosonic nonlocal field theory, examining correlation measures beyond entanglement entropy, namely, mutual information and tripartite information. Using numerical lattice simulations, we show that the nonlocality scale, \(A\), not only determines the onset of volume-law behavior but also leads to striking features: notably, extremely long-range mutual information and an unusual monogamy structure. In this regime, increasing the separation between large regions can paradoxically enhance their multipartite entanglement. Through holographic duality, we verify that the Ryu-Takayanagi formula correctly captures the volume-law scaling of entropy. Yet, a significant tension emerges: while the field theory reveals rich spatial correlations, the holographic model predicts a complete suppression of both mutual and tripartite information in the volume-law phase. This non-monogamous behavior in the holographic description stands in sharp contrast to the monogamous and highly structured entanglement observed in the field theory. Our results demonstrate that nonlocality gives rise to quantum states of such complexity that conventional geometric models of spacetime fall short. This points to the need for a new framework that goes beyond geometry to fully capture the nature of these correlations.	}

\begin{document} 
\maketitle
\flushbottom

\section{Introduction}\label{sec:intro}

Quantum entanglement is one of the most important differences between quantum and classical physics. It describes how parts of a system can be connected in ways that cannot be explained by any theory based on local hidden variables \cite{Bell:1964kc,Freedman:1972zza,Hensen:2015ccp}. Entanglement plays a key role in both the foundations of quantum theory \cite{Einstein:1935rr,Schrödinger:1935} and in practical areas like quantum information science. In the context of Quantum Field Theory (QFT), entanglement is a natural feature of the vacuum state. Although QFT respects locality through the microcausality condition, its states can still show strong nonlocal correlations. These correlations are often measured using entanglement entropy (EE). For a pure global state \( \rho \), dividing space into a region \( \Omega \) and its complement \( \bar{\Omega} \) gives a reduced density matrix \( \rho_\Omega = \text{Tr}_{\bar{\Omega}} \rho \) where  the von Neumann entropy,
\begin{equation}
S_\Omega = -\text{Tr}(\rho_\Omega \log \rho_\Omega),
\end{equation}
measures the entanglement between \( \Omega \) and \( \bar{\Omega} \).
In local QFTs, vacuum entanglement leads to a well-known result: the EE becomes infinite in the ultraviolet (UV) limit and follows an area law \cite{Bombelli:1986rw, Srednicki:1993im, Peschel:2002yqj,Eisert:2008ur}:
\begin{equation}
S_\Omega \sim \alpha \, \frac{\mathcal{A}(\partial \Omega)}{\epsilon^{d-2}} + \text{subleading terms},
\end{equation}
where \( \epsilon \) is a UV cutoff, \( \mathcal{A}(\partial \Omega) \) is the area of the boundary between regions, and \( \alpha \) depends on the theory. This area-law behavior is a clear sign of locality in quantum field theories. Although EE is a useful tool, it has some limitations. Its divergence in the UV limit makes it hard to compare across systems, and it includes all types of correlations without telling us what kind they are. To address these issues, mutual information is often used which measures the total correlations -both classical and quantum- between two separate regions \( \Omega_1 \) and \( \Omega_2 \). It is defined as:
\begin{equation}\label{mu}
I(\Omega_1,\Omega_2) = S_{\Omega_1} + S_{\Omega_2} - S_{\Omega_1 \cup \Omega_2}.
\end{equation}
Unlike EE, mutual information stays finite in the continuum limit because the area-law divergences cancel out \cite{Wolf:2007tdq}. The way mutual information decreases with distance helps us understand how entanglement spreads through the vacuum \cite{Groisman:2005dbo}. To explore entanglement beyond simple pairwise connections, the tripartite information is also used:
\begin{equation}\label{tri3}
I_3(\Omega_1,\Omega_2,\Omega_3) = I(\Omega_1,\Omega_2) + I(\Omega_1,\Omega_3) - I(\Omega_1,\Omega_2 \cup \Omega_3).
\end{equation}
This quantity tells us how much information about region \( \Omega_1 \) is shared jointly with regions \( \Omega_2 \) and \( \Omega_3 \). A key feature of \( I_3 \) is that it can be negative, which signals strong multipartite entanglement. The sign of \( I_3 \) helps us classify the type of correlation:
\begin{itemize}
	\item \( I_3 < 0 \):  This is a sign of genuine quantum entanglement shared among multiple parts \cite{Linden:2002vpn}. In standard quantum information theory, this is called monogamous because it indicates entanglement cannot be freely shared \cite{Hayden:2011ag, Koashi:2003pgf}. The total correlation between \( \Omega_1 \) and the combined system \( \Omega_2 \cup \Omega_3 \) is less than the sum of its pairwise correlations. 
	\item \( I_3 \geq 0 \): The correlation structure is more classical or mainly involves pairwise sharing.
\end{itemize}
This makes tripartite information and mutual information  a powerful diagnostic tool, and their true potential is realized in the context of the AdS/CFT correspondence. In this correspondence, the emergence of classical spacetime is linked to a specific quantum entanglement structure, where the Ryu-Takayanagi (RT) formula geometrizes entanglement entropy \cite{Ryu:2006bv}. Crucially, holographic states universally exhibit monogamous mutual information (\(I_3 < 0\)), making this a key diagnostic for a classical gravitational dual \cite{Hayden:2011ag, Mirabi:2016elb, Iizuka:2025ioc, Ju:2023tvo, Tanhayi:2017wcd,RezaMohammadiMozaffar:2016lbo,Tanhayi:2016uui}. This provides a clear mission: to test nonlocal field theories, which exhibit a radical volume-law entanglement entropy, against this geometric criterion. However, how nonlocality affects more detailed patterns of correlation, captured by measures like mutual information and tripartite information, and how these features might be represented in a holographic dual, remains largely unknown.

In this study, we present an analysis of entanglement in a bosonic nonlocal field theory. Using numerical lattice simulations, we calculate entanglement entropy, mutual information, and tripartite information to study the structure of correlations in the system. We also perform holographic calculations based on a proposed gravitational dual to compare with the field theory results. Our findings confirm that strong nonlocality leads to volume-law scaling of entanglement entropy. However, we uncover a key tension: although the holographic dual correctly reproduces the volume-law behavior, it fails to reflect the rich and monogamous patterns of bipartite and multipartite correlations seen in the field theory. This mismatch highlights a limitation of classical spacetime geometry in fully capturing the complex entanglement structure of strongly nonlocal quantum states.

The structure of this work is as follows. Section 2 provides a brief overview of the bosonic nonlocal field theory and outlines the numerical lattice methods used to compute entanglement measures. We present results for entanglement entropy, mutual information, and tripartite information, showing how the nonlocality scale controls the shift from area-law to volume-law behavior and reshapes the pattern of correlations. Section 3 focuses on the holographic dual of the theory. We derive the geometric expression for entanglement entropy using the RT formula and examine the holographic predictions for mutual and tripartite information. These results are compared directly with the field theory calculations. Finally, Section 4 summarizes our findings, discusses the implications of the mismatch between the two approaches, and offers concluding remarks.

\section{  Entanglement Structure of Nonlocal Fields}

To systematically investigate how nonlocality modifies quantum entanglement, we employ a lattice regularization that permits exact computation of entanglement measures. We model a real scalar field as a network of coupled quantum harmonic oscillators on a spatial lattice, with sites labeled by capital Latin indices \( M, N \). This discretization serves both as a UV regulator and as an analytically tractable framework for entanglement computation \cite{Srednicki:1993im,Peschel:2002yqj,Shiba:2013jja}.\footnote{For further studies and related computational approaches, see \cite{MohammadiMozaffar:2024uiy,MohammadiMozaffar:2017nri,Vasli:2023syq, Doroudiani:2019llj,Khorasani:2023usq, Casini:2005zv, Azeyanagi:2007bj, Herzog:2012bw, Herzog:2013py, Pirmoradian:2021wvo,Pirmoradian:2023uvt, Pirmoradian:2025dco}.}
The system dynamics are governed by the Hamiltonian:
\begin{equation}
	\label{eq:hamiltonian}
	H = \frac{1}{2} \delta_{M N} P^{M}P^{N} + \frac{1}{2}V_{M N} q^{M} q^{N},
\end{equation}
where \( q_{M} \) and \( P_{M} \) represent the displacement and momentum of the \( M \)-th oscillator, respectively. The symmetric, positive-definite coupling matrix \( V_{MN} \) encodes all interactions between lattice sites and is independent of the dynamical variables.
For this Gaussian system, the ground state wavefunction admits an exact expression:
\begin{equation}
	\label{eq:wavefunction}
	\psi(\{q_{M}\}) = \left(\det \dfrac{W}{\pi} \right)^{1/4} \exp\left(-\frac{1}{2}W_{MN}q^{M}q^{N} \right),
\end{equation}
where the matrix \( W \equiv V^{1/2} \) is the positive square root of the coupling matrix. This Gaussian structure enables efficient computation of entanglement entropy via covariance matrix techniques \cite{Peschel:2002yqj}.
To compute the entanglement entropy of a spatial subregion \( \Omega \), we partition the lattice into \( \Omega \) and its complement \( \bar{\Omega} \), adopting the convention where Latin indices \( (a, b) \) denote sites in \( \Omega \) and Greek indices \( (\alpha, \beta) \) represent sites in \( \bar{\Omega} \). The \( W \) matrix and its inverse are correspondingly partitioned as:
\begin{equation}
	\label{eq:block_partition}
	W_{MN} = \begin{pmatrix}
		W_{ab} & W_{a\beta} \\ W_{\alpha b} & W_{\alpha\beta}
	\end{pmatrix} \equiv 
	\begin{pmatrix}
		K & L \\ L^{T} & M
	\end{pmatrix}, \quad
	W^{-1}_{MN} = \begin{pmatrix}
		\widetilde{W}_{ab} & \widetilde{W}_{a\beta} \\ \widetilde{W}_{\alpha b} & \widetilde{W}_{\alpha\beta}
	\end{pmatrix} \equiv 
	\begin{pmatrix}
		P & Q \\ Q^{T} & R
	\end{pmatrix}.
\end{equation}
The entanglement entropy of region \( \Omega \) is then given by:
\begin{equation}
	S_{\Omega} = \sum_{n} \Big[
	\left(\frac{\lambda_n + 1}{2} \right) \ln \left(\frac{\lambda_n + 1}{2} \right) - \left(\frac{\lambda_n - 1}{2} \right) \ln \left(\frac{\lambda_n - 1}{2} \right)\Big],	\label{eq:f_lambda}
\end{equation}
where \( \lambda_{n} \) are the eigenvalues of the following matrix:
\begin{equation}
	\label{eq:lambda_matrix}
	\Lambda^{a}_{\, b} = (W^{-1})_{ac} (W)^{cb} = P_{ac} K^{cb}.
\end{equation}
The matrices \( P \) and \( K \) possess clear physical interpretations: \( K \equiv W_{ab} \) encodes direct couplings between oscillators within \( \Omega \), while \( P \equiv (W^{-1})_{ab} \) governs the two-point correlation functions \( \langle q_a q_b \rangle \). The product matrix \( \Lambda = P \cdot K \) thus represents a composite operator mixing coupling and correlation information within the subregion. Each eigenvalue \( \lambda_n > 1 \) corresponds to an entangled mode between \( \Omega \) and the remainder of the system, with the summation in Eq. (\ref{eq:f_lambda}) quantifying the total entanglement entropy.
We first apply this formalism to a conventional free scalar field in 2-dimensional Minkowski spacetime, discretized with lattice spacing \( a \) and periodic boundary conditions. The dimensionless Hamiltonian becomes:
\begin{equation}
	\label{eq:hamiltonian_discrete}
	H_{0} \equiv aH = \sum_{n} \frac{1}{2}\pi^{2}_{n} + \sum_{m,n} \frac{1}{2}\phi_{m} V_{mn} \phi_{n},
\end{equation}
which converges to the continuum field theory as \( N \to \infty \). This local framework provides the foundation for introducing nonlocal interactions.

\subsection{Generalization to Nonlocal Scalar Field Theory}

The preceding analysis of local field theories can be extended to models featuring nonlocal spatial interactions. This generalization is motivated by the profound impact nonlocality has on quantum entanglement. While local theories with short-range interactions are characterized by an area-law entanglement entropy, nonlocal couplings can facilitate a volume-law scaling, where the entropy grows proportionally with the volume of the subsystem \cite{Srednicki:1993im, Koffel:2012cu}. To investigate this, we consider a massive scalar field theory where the locality of the kinetic term is relaxed. The system is described by the Hamiltonian:
\begin{equation}
	\label{eq:hamiltonian_nonlocal}
	H = \frac{1}{2} \int dx \left[ \left( \dfrac{\partial \phi}{\partial t} \right)^{2} + B_{0} \, \phi \, e^{A_{0}(- \partial^{2})^{\omega/2}} \phi \right].
\end{equation}
Here, the nonlocal operator \( e^{A_{0}(- \partial^{2})^{\omega/2}} \) governs the field's spatial interactions. The key parameters are:
\( A_0 \): The nonlocality scale, with dimensions of $(length)^\omega$.  
	The nonlocality exponent \( \omega \) determines the asymptotic behavior of the interaction.
Moreover, \( B_0 \) stands for  a mass-scale parameter.

To connect the continuum theory to a lattice regulator, a lattice spacing \( a \) is introduced. Thus the parameters scale as:
\begin{equation}
	\label{eq:parameter_scaling}
	B_{0} = \frac{B}{a^{2}}, \qquad A_{0} = a^{\omega} A,
\end{equation}
where \( A \) and \( B \) are the dimensionless lattice parameters. Notably, in lattice units \( B \) can be rescaled to unity without loss of generality, as it does not alter the entanglement properties of the ground state.

The entanglement entropy for a subsystem on the lattice is determined by the matrices \( W^\omega \) and their inverses, which are defined via their Fourier transforms:

\begin{equation}
	\label{eq:W_lattice}
	\begin{split}
		(W^{\omega})_{mn} &= \int^{\pi}_{-\pi} \dfrac{dq}{2\pi} \, e^{i q(n-m)} \, \exp \left[ \frac{A}{2} \left( 2 - 2 \cos q \right)^{\omega/2} \right], \\
		(W^{\omega})^{-1}_{mn} &= \int^{\pi}_{-\pi} \dfrac{dq}{2\pi} \, e^{i q(n-m)} \, \exp \left[ -\frac{A}{2} \left( 2 - 2 \cos q \right)^{\omega/2} \right].
	\end{split}
\end{equation}
For general \( \omega \), the evaluation of these integrals leads to expressions involving generalized hypergeometric functions. In particular, for even values of \( \omega \), the matrix elements \( (W^{\omega})_{nm} \) can be expressed as a series:
\begin{multline}\label{hyper}
	(W^{\omega})_{nm} = \, _{\frac{\omega}{2}}F_{\frac{\omega}{2}}\left(\left\{\frac{2i-1}{\omega}\right\}_{i=1}^{\omega/2}; \left\{\frac{2i}{\omega}\right\}_{i=1}^{\omega/2}; 2^{\omega-1} A\right) + \\
	\sum_{i=1}^{\infty} (-1)^{i} \mathcal{C}_i(n-m)  \, _{\frac{\omega}{2}}F_{\frac{\omega}{2}}\left(\left\{\frac{2(i+j)-1}{\omega}\right\}_{j=1}^{\omega/2}; \left\{\frac{2(i+j)}{\omega}\right\}_{j=1}^{\omega/2}; 2^{\omega-1} A\right),
\end{multline}
where \( \mathcal{C}_i \) are integer coefficients. While this formal solution is instructive, its complexity motivates a focus on specific, analytically tractable cases.

For the sake of clarity and to obtain closed-form results, this study will concentrate on the exponents \( \omega = 1 \) and \( \omega = 2 \). In these cases, the expressions simplify significantly to well-known special functions:

\begin{eqnarray}
	(W^{1})_{nm} &=& J_{2|n-m|}(i A)+i\,\pmb{E}_{2|n-m|}(i A),
	\label{Eq:weqal1}\\	
	(W^{2})_{nm} &=& (-1)^{|n-m|}e^A I_{|n-m|}(A). 
	\label{Eq:weqal2}
\end{eqnarray}
Here, \( J_{\nu} \) is the Bessel function of the first kind, \( \pmb{E}_{\nu} \) is the Weber function, and \( I_{\nu} \) is the modified Bessel function of the first kind. The corresponding expressions for the inverse matrix \( (W^{\omega})^{-1}_{nm} \) are obtained by substituting \( A \rightarrow -A \) in the equations above.

\begin{figure}[tbp]
\centering % \begin{center}/\end{center} takes some additional vertical space
\includegraphics[width=.496\textwidth,origin=c]{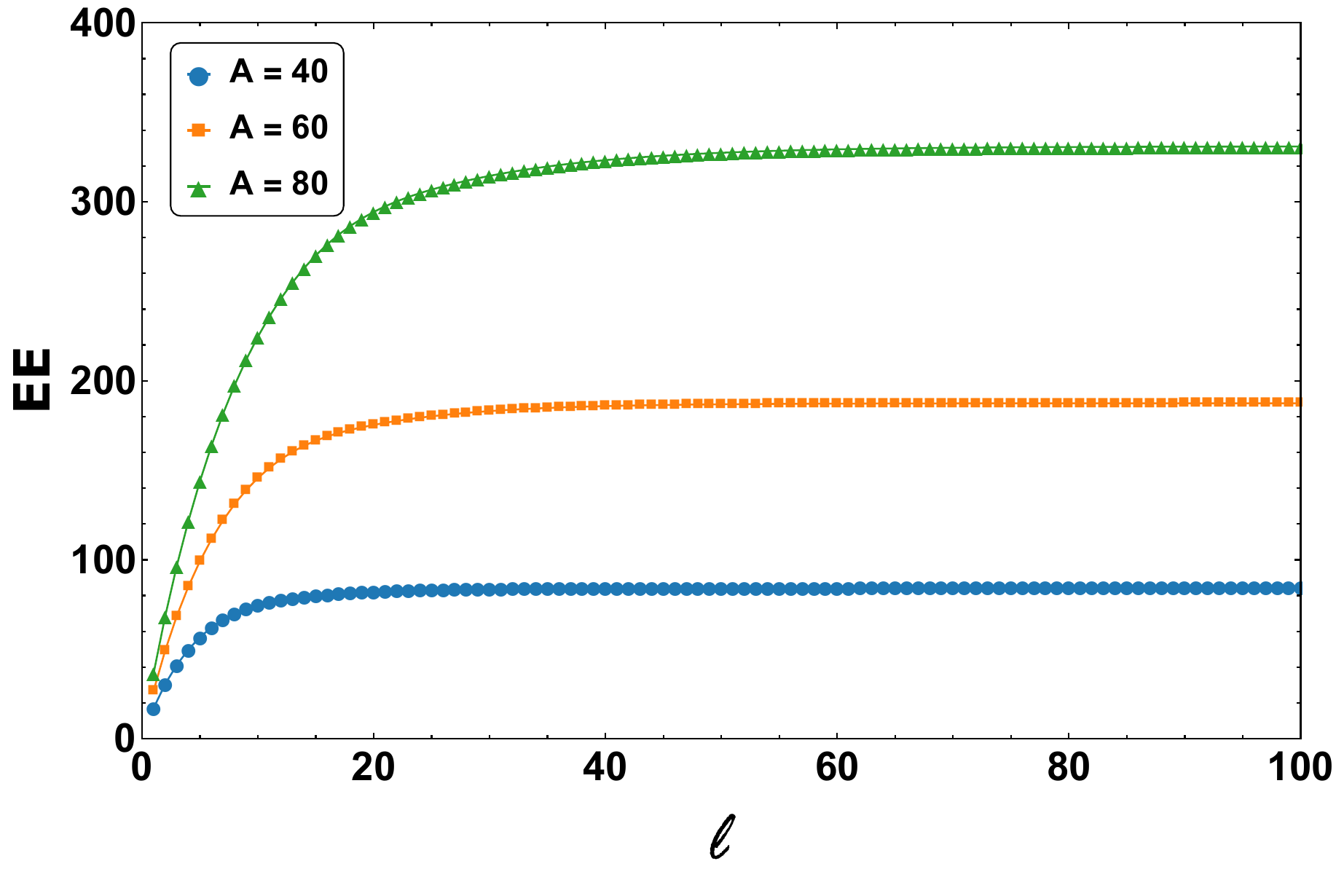}
\includegraphics[width=.496\textwidth,origin=c]{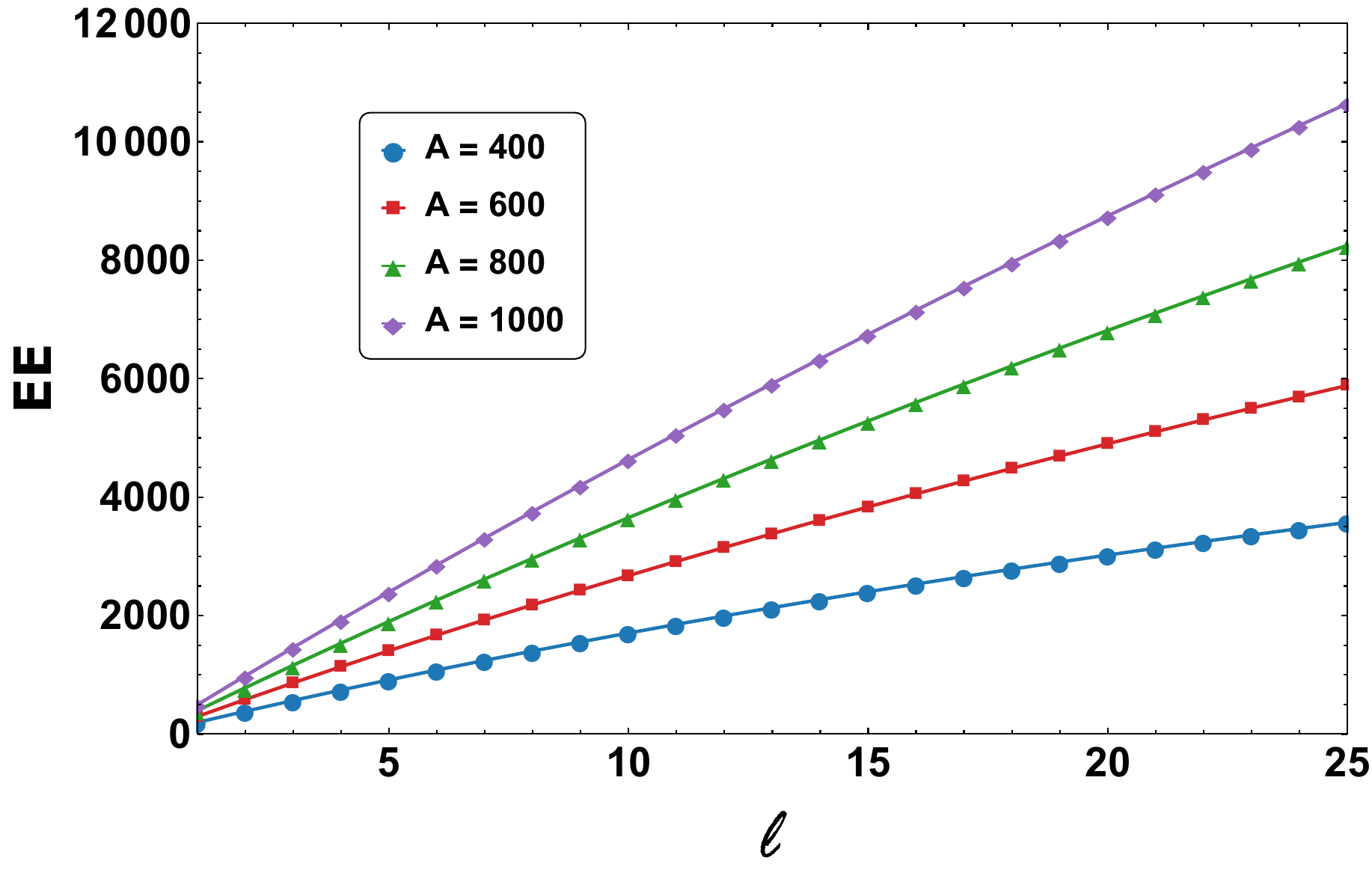}
\caption{Entanglement entropy \( S_\Omega \) as a function of subsystem length \( l \) for varying nonlocality parameters \( A \). Left: For moderate nonlocality (\( A = 40, 60, 80 \)), the entropy exhibits a volume-law scaling regime for small \( l \), with the extent of this regime expanding as \( A \) increases. Right: For large nonlocality (\( A = 400, 600, 800, 1000 \)), the volume-law scaling is significantly enhanced and persists over a much wider range of \( l \), demonstrating the profound impact of strong nonlocality on the entanglement structure. \label{fig:2.1}}
\end{figure}

Our numerical computation of entanglement entropy proceeds by constructing  correlation matrix \( \Lambda \) via Eq. (\ref{eq:lambda_matrix}) and diagonalize it to obtain the eigenvalues \( \lambda_n \). Figures \ref{fig:2.1}  display the resulting entanglement entropy as a function of subregion length $l$ for parameter values $A = 40, 60, 80$ and $A = 400, 600, 800, 1000$. These results confirm established behavior for this model \cite{Shiba:2013jja} and provide a foundation for our original investigation of mutual information. The analysis reveals that the nonlocality parameter $A$ fundamentally controls the theory's entanglement structure. For subsystems  \( l \ll A \), the entropy exhibits clear volume-law scaling \( S \sim l \), directly contrasting with the area-law behavior of local quantum field theories. Crucially, the spatial extent of this volume-law regime expands systematically with increasing $A$, as evidenced in the right panel of Figure \ref{fig:2.1} where linear entropy growth persists across significantly larger $l$ values for $A = 400-1000$. This establishes that enhancing nonlocality not only increases overall entanglement but also extends the spatial domain over which the vacuum state exhibits extensive volume-law entanglement, confirming $A$'s role as the central scale governing the transition between nonlocal and effectively local behavior.

%\begin{figure}[tbp]\centering\includegraphics[width=1.0\textwidth,origin=c]{EE400,600,800,1000.pdf}
% "\includegraphics" is very powerful; the graphicx package is already loaded\caption{\label{fig:2.2} Entanglement entropy \( S_\Omega \) versus subsystem length \( L \) for large nonlocality parameters \( A = 400, 600, 800, 1000 \). The volume-law scaling is significantly enhanced and persists over a wider range of \( L \), demonstrating the profound impact of strong nonlocality on the entanglement structure.}\end{figure}

%\begin{table}[tbp]
%\centering
%\begin{tabular}{|lr|c|}
%\hline
%x&y&x and y\\
%\hline 
%a & b & a and b\\
%1 & 2 & 1 and 2\\
%$\alpha$ & $\beta$ & $\alpha$ and $\beta$\\
%\hline
%\end{tabular}
%\caption{\label{tab:i} We prefer to have borders around the tables.}
%\end{table}

\subsection{Mutual Information and Nonlocal Correlations}

The total correlations, both classical and quantum contributions, between two disjoint spatial regions \(\Omega_1\) and \(\Omega_2\) are quantified by their mutual information. A key advantage of mutual information is its finiteness in the continuum limit, in contrast to the ultraviolet-divergent EE, making it a robust and regulator-independent measure. % It is defined as
%\begin{equation} 	I(\Omega_1, \Omega_2) = S(\Omega_1) + S(\Omega_2) - S(\Omega_1 \cup \Omega_2),\end{equation} where \(S(\Omega)\) is the entanglement entropy of region \(\Omega\).
mutual information is non-negative and vanishes if and only if the reduced density matrices factorize, \(\rho_{\Omega_1 \cup \Omega_2} = \rho_{\Omega_1} \otimes \rho_{\Omega_2}\), indicating a complete absence of correlations. It also obeys strong subadditivity,
\begin{equation}
I(\Omega_1; \Omega_2 \cup \Omega_3) \geq I(\Omega_1; \Omega_2) + I(\Omega_1; \Omega_3) - I(\Omega_2; \Omega_3),
\end{equation}
which constrains the distribution of information among multiple subsystems. The behavior of mutual information serves as a sensitive probe of locality. This inequality constrains how information can be distributed among multiple subsystems and is closely related to the non-positivity of tripartite information in holographic systems.  In local QFTs, mutual information between spatially separated regions decays rapidly with distance, consistent with short-range interactions. In contrast, our nonlocal scalar field model features direct couplings between distant lattice sites, governed by the parameter \(A\). We therefore expect that increasing \(A\) systematically amplifies both the magnitude and the spatial range of the mutual information.
The behavior of mutual information provides a definitive signature of the profound structural changes induced by nonlocality, setting it in stark contrast to local quantum field theories where correlations decay rapidly with distance. Our results demonstrate that the nonlocality parameter \(A\) systematically reshapes the correlation architecture, generating entanglement networks that are both stronger and more extensive.

\begin{figure}[tbp]
\centering
\includegraphics[width=.495\textwidth]{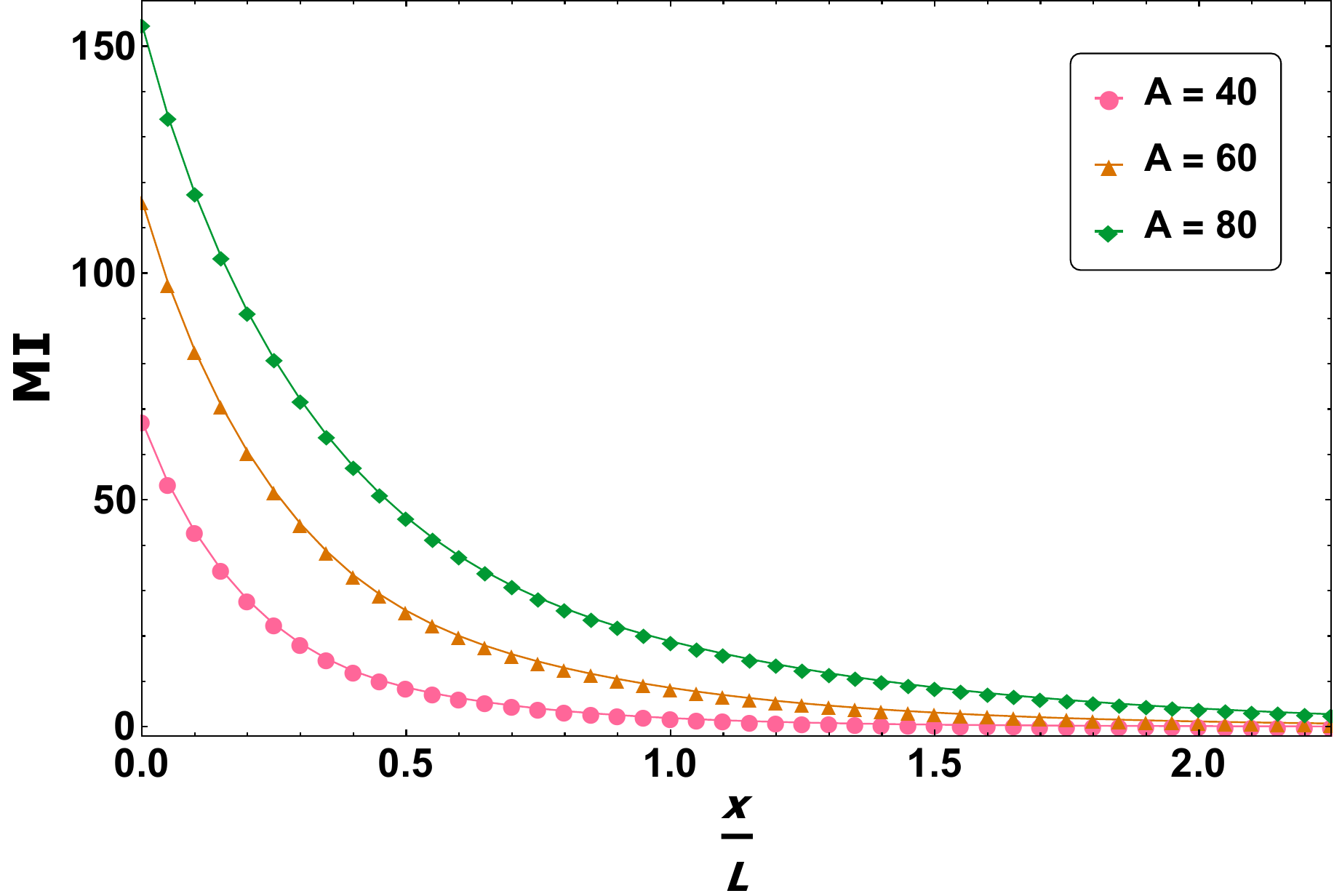}
\includegraphics[width=.495\textwidth]{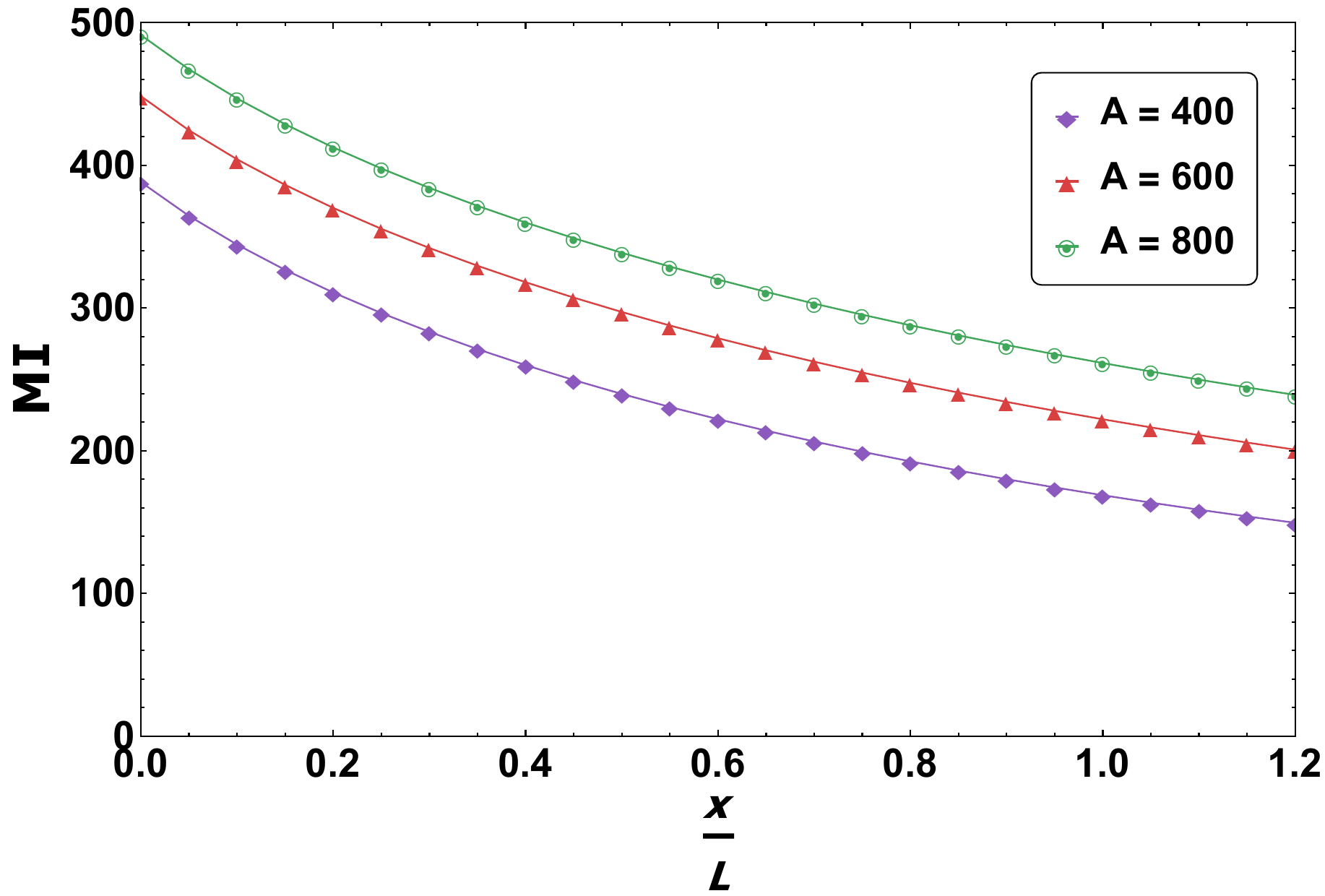}
\caption{Mutual information between two disjoint intervals of equal length (\( l_1 = l_2 = 10 \)) as a function of the normalized separation \( x/L \) where $L=l_{1}+ l_{2}$ is the total length of subsystems. Left: For moderate nonlocality parameters (\( A = 40, 60, 80 \)), the mutual information exhibits a gradual decay with separation, highlighting the persistence of correlations at large distances. Right: For strong nonlocality (\( A = 400, 600, 800 \)), the mutual information remains substantial across a wide range of separations, confirming that strong nonlocality drastically slows the decay of quantum correlations. \label{fi3}}
\end{figure}

A direct comparison between the left and right panels of Figure \ref{fi3}, showing moderate nonlocality ($A=40, 60, 80$) and strong nonlocality ($A=400, 600, 800$) respectively, reveals that the parameter $A$ acts as a master control for the range of quantum correlations, systematically stretching out their decay with distance. The left panel shows a gradual decline in mutual information, confirming the emergence of longer-range correlations, while the right panel demonstrates that for large $A$ this decay becomes exceptionally slow, with correlations remaining substantial even at significant separations. This powerful contrast confirms that increasing the nonlocality parameter $A$ not only enhances the strength of the quantum connections but, more importantly, fundamentally extends their spatial reach, transforming the vacuum into a state with pervasive, long-range entanglement.

\subsection{Tripartite information}
Another key quantity for understanding multipartite correlations is the tripartite information. This measure becomes relevant when a region \( \Omega \) is divided into three non-overlapping subregions: \( \Omega_1 \), \( \Omega_2 \), and \( \Omega_3 \). It is defined by \eqref{tri3} and can be written as:
\begin{equation}
\label{eq:4.1}
I_{3}(\Omega_{1},\Omega_{2},\Omega_{3})= S_{\Omega_{1}} + S_{\Omega_{2}}+S_{\Omega_{3}}- S_{\Omega_{1} \cup \Omega_{2} }
- S_{\Omega_{1} \cup \Omega_{3} }- S_{\Omega_{2} \cup \Omega_{3} }
 +  S_{\Omega_{1} \cup \Omega_{2}\cup \Omega_{3} }
\end{equation}
The behavior of \( I_3 \) reveals a clear signature of a significant transition in the quantum vacuum, driven by the nonlocality parameter \( A \). This transition is evident when comparing different regimes. For small \( A \), \( I_3 \) remains close to zero, indicating a quasi-local state where correlations are mostly bipartite and can be shared. As \( A \) increases, however, the system undergoes a major reorganization. Strongly negative values of \( I_3 \) emerge, signaling a shift to a regime dominated by the monogamy of entanglement, where correlations are genuinely multipartite and cannot be freely shared. Remarkably, in this strongly nonlocal regime, increasing the distance between large subsystems makes \( I_3 \) even more negative. This means that greater separation actually strengthens their collective quantum entanglement, a counterintuitive result. These findings show that the nonlocality scale \( A \) not only enhances entanglement but also reshapes its structure, creating a complex correlations across distant regions. This represents a significant departure from the behavior expected in local quantum field theories.
\begin{figure}[tbp]
	\centering
	\includegraphics[width=.495\textwidth,origin=c]{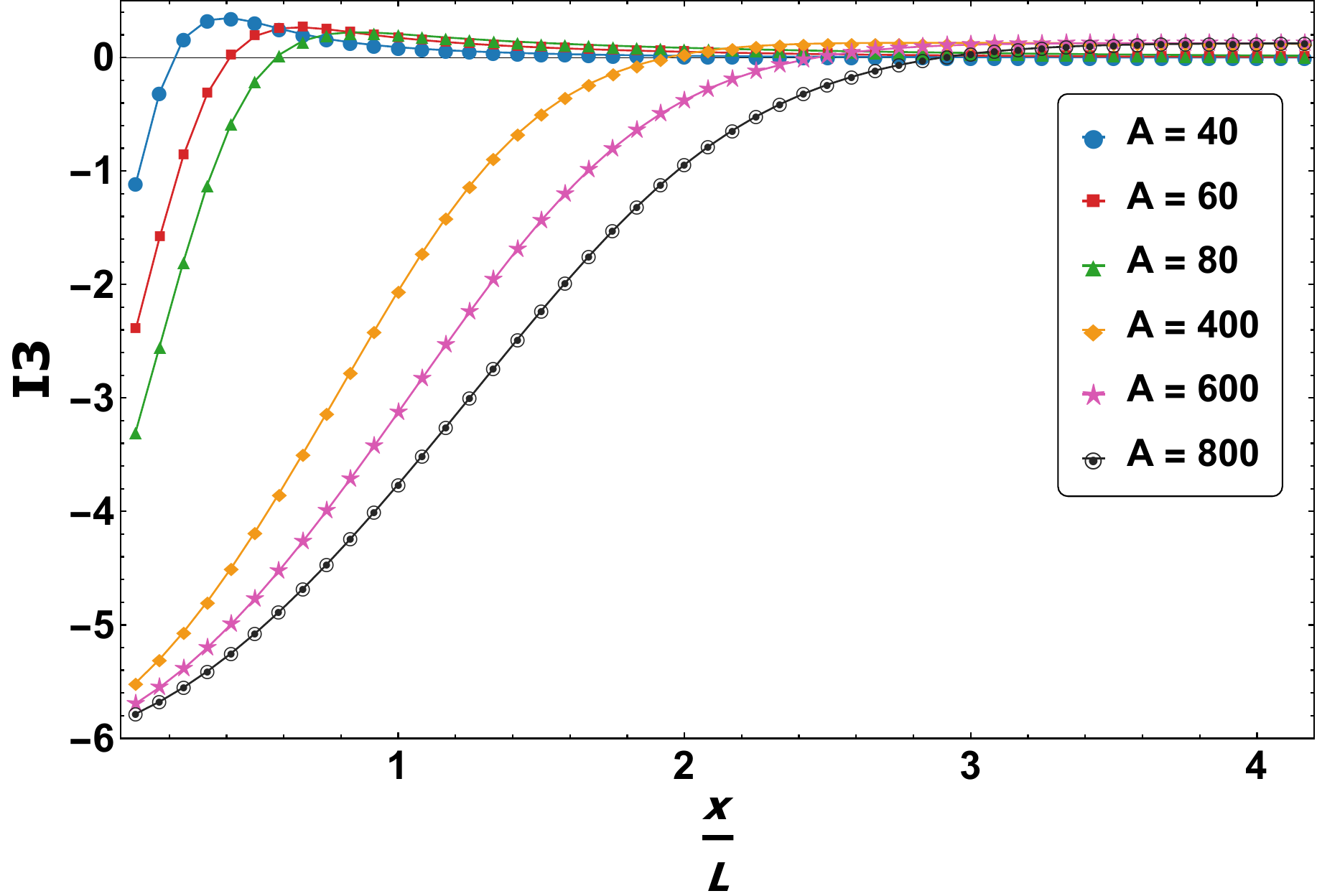}
	\includegraphics[width=.495\textwidth,origin=c]{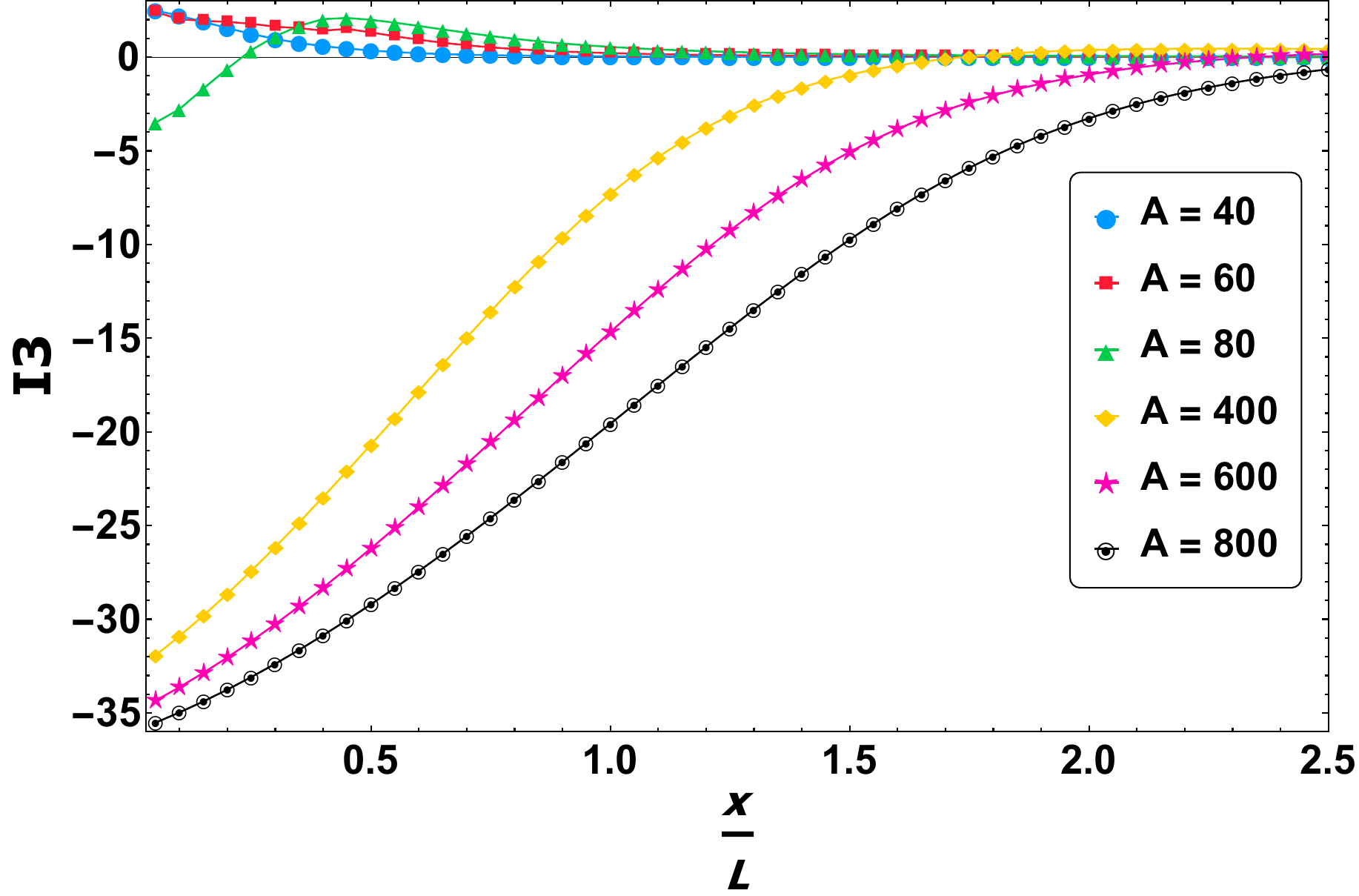}
	\caption{Tripartite information \( I_3 \) for three disjoint regions as a function of their separation. Left: For small, equal-length regions, the tripartite information is consistently negative, confirming the monogamous nature of the entanglement. Right: For larger, equal-length regions, the negative values become significantly more pronounced with separation, especially for higher \( A \).  \label{fig:tripartite}}
\end{figure}
%\begin{figure}[tbp]	\centering	\includegraphics[width=1.0\textwidth,origin=c]{TI L=10.pdf}
	
	% \includegraphics[width=1.0\textwidth,origin=c]{HolographyTP L=30}	\caption{Tripartite information for three larger, equal-length regions. Here, the negative values become more pronounced as the regions are moved apart, especially for higher values of $A$. This reveals that larger regions can access the long-range "quantum network" created by nonlocality, leading to stronger and more complex multipartite entanglement over greater distances.}\label{fig8}\end{figure}
Figure \ref{fig:tripartite} demonstrates that the impact of nonlocality on multipartite entanglement is governed by the subsystem size. For small regions, entanglement is dominated by local interactions and is largely insensitive to the nonlocality parameter \( A \). For larger regions, the long-range correlations induced by a high \( A \) become dominant, enabling a much more robust and widely distributed multipartite entanglement structure.
%%%%%%%%%%%%%%%%%%%%%%%%%%%%%%%%%%%%%%%%%%%%%%%%%%%%%
\section{Holographic Entanglement and the Geometry of Nonlocality}
To investigate the gravitational dual of the nonlocal field theory, we build upon the holographic framework established by Shiba and Takayanagi \cite{Shiba:2013jja}. The cornerstone of this approach is the RT prescription \cite{Ryu:2006bv}, which states that the entanglement entropy of a boundary region \(\Omega\) is given by the area of the minimal bulk surface \(\gamma_{\Omega}\) anchored to its boundary:
\[
S_{\Omega} = \frac{\text{Area}(\gamma_{\Omega})}{4G_{N}}.
\]
This foundational result provides a geometric method for computing entanglement entropy. Our analysis begins by recalling this holographic entanglement entropy (HEE) for our nonlocal theory, confirming how nonlocality distorts the bulk geometry and modifies the RT surfaces. 

However, the central objective of this work is to extend this holographic analysis beyond entanglement entropy. We employ more correlation measures, mutual information and tripartite information, to probe the finer, non-local structure of the quantum vacuum that entropy alone cannot reveal.

The boundary nonlocality, introduced via the operator \(e^{A_0 (-\partial^2)^{\omega/2}}\), must have a geometric imprint in the bulk. This imprint is captured by a warping of the metric along the holographic radial direction. As established in \cite{Shiba:2013jja,Nozaki:2012zj}, the crucial metric component is:
\begin{equation}
	g_{uu} = \frac{A_{0}^{2}}{a^{2}}e^{2\omega u},
\end{equation}
where \(u\) is the original radial coordinate. Employing the standard coordinate transformation \(z = a e^{-u}\) (placing the boundary at \(z=a\)), the full bulk metric becomes:
\begin{equation}
	ds^{2} \propto A_{0}^{2} \frac{dz^{2}}{z^{2(\omega+1)}} + \frac{1}{z^{2}}\sum_{i=1}^{d-1}dx_{i}^{2} + g_{tt}dt^{2}.
\end{equation}
Using \(y = z^{-\omega}\) further simplifies the spatial components, resulting in a form that is more suitable for calculating minimal surfaces:
\begin{equation}
	ds^{2} \propto A^{2}_{0} dy^{2} + y^{\frac{2}{\omega}} \sum^{d-1}_{i=1} dx^{2}_{i}.
\end{equation}
We consider a boundary subregion \(\Omega\) given by a strip of width \(l\). The area of a bulk surface \(\gamma_{\Omega}\) anchored to \(\partial \Omega\) is described by the functional:
\begin{equation}
	\text{Area} = R^{d-2} a^{d-2} \int^{l/2}_{-l/2} dx_{1} \ y^{\frac{d-2}{\omega}} \sqrt{A^{2}_{0} (y')^{2} + y^{\frac{2}{\omega}}}.
\end{equation}
Varying this functional yields the equation for the minimal surface, which extends from a deepest point \(y_*\) to a UV cutoff \(y_{\infty}=a^{-\omega}\). This cutoff is the bulk dual to the boundary cutoff in the field theory, giving the range \(y_{*} \leqslant y < y_{\infty}\). The corresponding area can be expressed as:
\begin{equation}\label{eq:6.13}
	\text{Area}= 2R^{d-2}a^{d-2} A_{0} \int^{y_{\infty}}_{y_{*}} dy\ \dfrac{y^{(2d-3)/\omega}\,y_{*}^{-(d-1)/\omega}}{\sqrt{y^{2(d-1)/\omega}\,\,y_{*}^{-2(d-1)/\omega}-1}}.
\end{equation}
The problem simplifies significantly for a 2-dimensional boundary (\(d=2\)) with a nonlocality exponent \(\omega=1\), due to a conservation law. In this case, the minimal surface profile \(y(x)\) is determined by:
\begin{equation}
	A_{0} \frac{dy}{dx_{1}} = y \sqrt{ \frac{y^{2}}{y_{*}^{2}} - 1 },
\end{equation}
where \(y_{*}\) is an integration constant identified as the turning point of the surface, defined by \(dy/dx_{1}=0\). Consequently, the area simplifies to:
\begin{equation}
	\text{Area} = 2 A_{0} \int^{y_{\infty}}_{y_{*}} dy\ \frac{y\ y_{*}^{-1}}{\sqrt{(y^{2}/y_{*}^{2})-1}}.
\end{equation}
Our analysis reveals two distinct entanglement regimes, governed by the ratio of the subsystem size \(l\) to the nonlocality scale \(A\):

\subsubsection*{Volume-Law Regime (\(l \ll A\))}
When the size of the probed region is smaller than the nonlocality scale (\(l \ll A\)), the connected minimal surface remains near the boundary, with its turning point approaching the UV cutoff (\(y_{*} \simeq y_{\infty} = a^{-1}\)). In this regime, the area exhibits extensive, volume-law scaling with the subsystem size, \(\text{Area} \propto l\). This linear dependence is the holographic signature of the highly entangled vacuum state in a nonlocal theory. As shown in Figure \ref{fig:6.1}, the slope of this linear growth increases with the nonlocality scale \(A\), indicating that stronger nonlocality enhances the entanglement density at short distances.

\subsubsection*{Saturation Regime (\(l \gg A\))}
For large subsystem sizes, the connected minimal surface is no longer the dominant saddle point. The entanglement entropy is instead given by the sum of two disconnected surfaces anchored at the strip's boundaries:
\begin{equation}
	x_1 = \pm \frac{l}{2}, \qquad 0 < y < y_{\infty}.
\end{equation}
This geometric transition implies that the entropy saturates to a constant value, \(S \propto A\), independent of \(l\). This establishes \(A\) as the fundamental scale that caps the amount of retrievable entanglement from a spatial region.

\begin{figure}[tbp]
\centering
\includegraphics[width=.495\textwidth,origin=c]{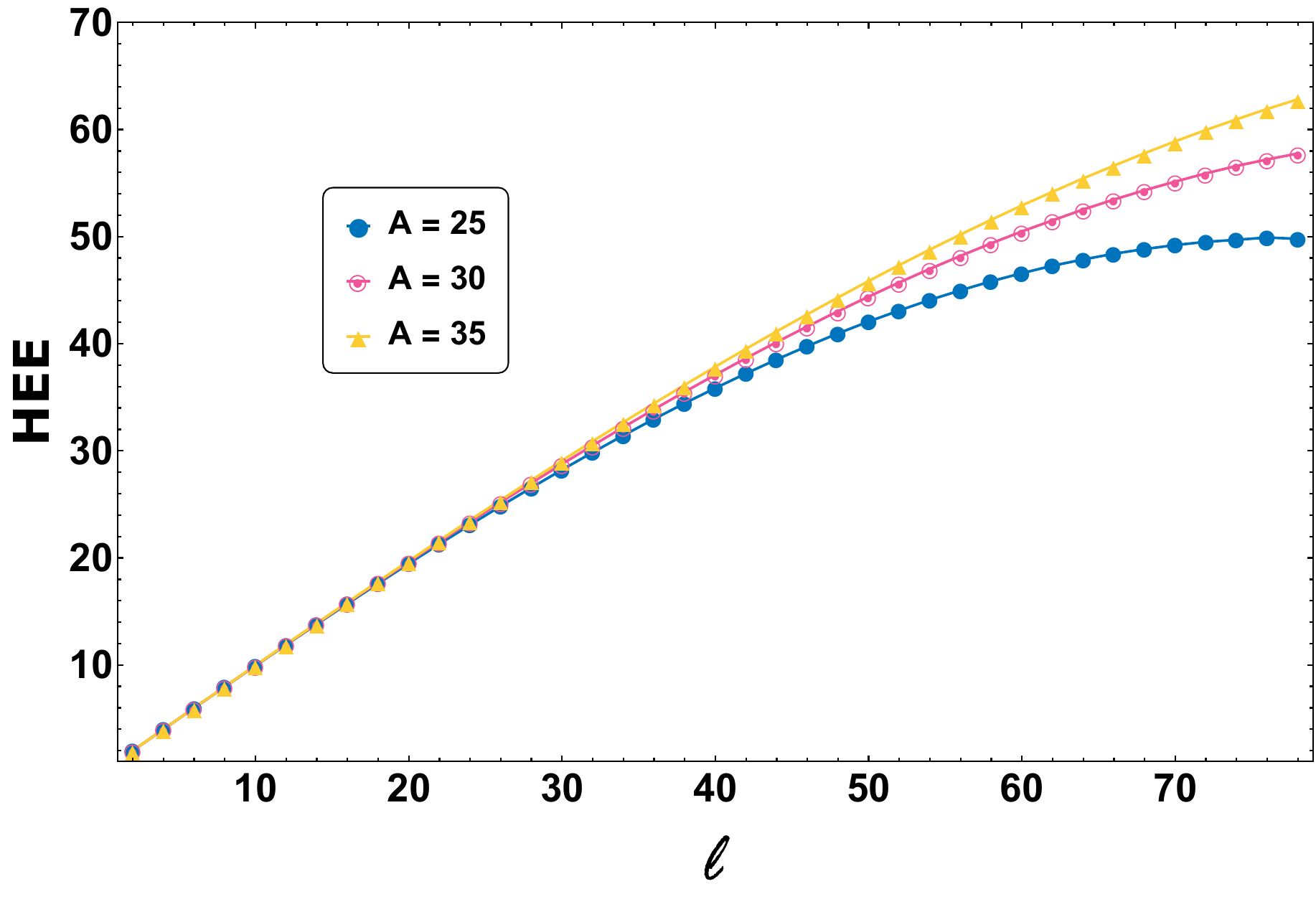}
\includegraphics[width=.495\textwidth,origin=c]{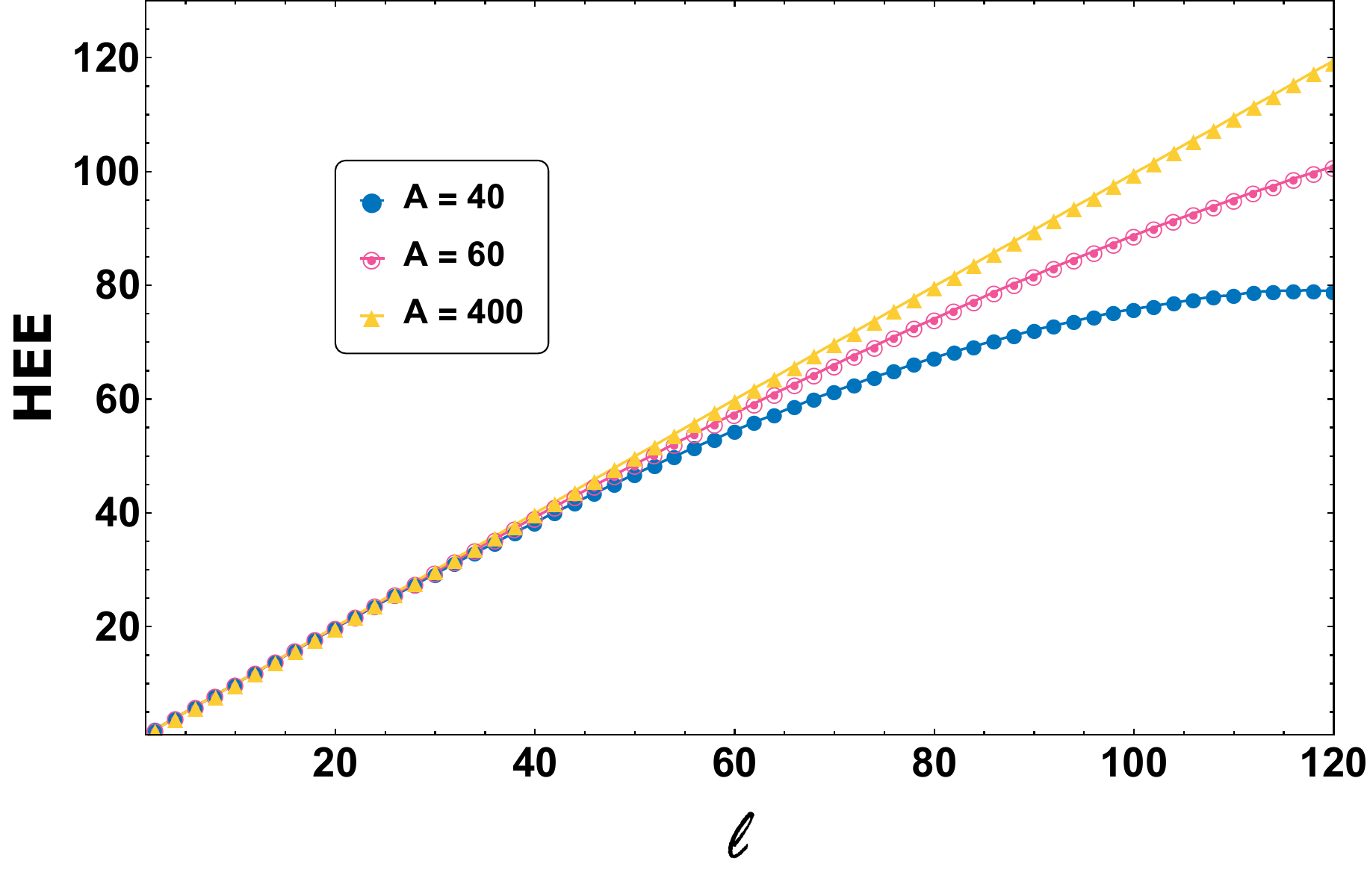}
\caption{Holographically computed minimal surface area (proportional to entanglement entropy) for \( d=2 \), \( \omega=1 \) as a function of subsystem size \( l \). Left: For small nonlocality parameters (\( A = 25, 30, 35 \)), the area exhibits a clear linear scaling with \( l \) in the regime \( l \ll A \), confirming a volume law. Right: For a broader range of parameters from moderate to large (\( A = 40, 60, 400 \)), the volume-law scaling persists, and the slope of the linear growth is significantly enhanced with increasing \( A \), demonstrating its role in amplifying entanglement. \label{fig:6.1}
 }
\end{figure}

Our holographic results for entanglement entropy confirm the expected pattern: the nonlocality parameter \(A\) warps space, creating extensive entanglement (volume law) for small regions and a saturation of entropy for large ones, as seen in Figure \ref{fig:6.1}. However, the main achievement of our work is the use of more powerful tools: mutual information and tripartite information, to investigate the nature of these quantum connections. Mutual information shows that the links between separate regions fade very slowly, extending far beyond what is possible in ordinary theories. The consistently negative tripartite information proves this entanglement is complex and shared between multiple parties, not just simple pairs.

%\begin{figure}[tbp]\centering\includegraphics[width=1.0\textwidth,origin=c]{HEE25,30,35.pdf}\caption{\label{fig:6.1}Holographically computed minimal surface area (proportional to entanglement entropy) for \( d=2 \), \( \omega=1 \) as a function of \( l \), for \( A = 25, 30, 35 \). In the regime \( l \ll A \), the area scales linearly with \( l \), confirming a volume law. The slope of this linear growth increases with \( A \). }\end{figure}

\subsection{ Holographic Mutual Information }
To probe the spatial structure of correlations in our nonlocal theory, we compute the holographic mutual information between two disjoint boundary regions \(\Omega_{1}\) and \(\Omega_{2}\) defined by \eqref{mu}. In holographic setup, calculating 	$S_{\Omega_{1} \cup \Omega_{2}}$ involves a geometric competition in the bulk between two candidate RT surfaces:   A disconnected configuration, comprising the separate minimal surfaces for two regions.  A connected configuration, where a single surface bridges two regions in the bulk. The holographic prescription selects the configuration with the minimal area:
\begin{equation}
	S_{\Omega_{1} \cup \Omega_{2}} = \frac{1}{4G_N} \min \left\{ \mathcal{A}_{\text{dis}}, \, \mathcal{A}_{\text{con}} \right\}.
\end{equation}
Consequently, the mutual information undergoes a sharp transition:
\begin{equation}
	I(\Omega_{1},\Omega_{2}) =
	\begin{cases}
		\dfrac{1}{4G_N} \left( \mathcal{A}_{\text{dis}} - \mathcal{A}_{\text{con}} \right), & \text{connected phase}, \\[1em]
		0, & \text{disconnected phase}.
	\end{cases}
\end{equation}
A non-zero holographic mutual information signals the dominance of a connected bulk surface, implying robust bipartite correlations between boundary regions. Our key finding is that the nonlocality parameter \(A\) acts as a master switch for this geometric transition. For a fixed subsystem size (\(l=50\)), Figure \ref{fig:7.1} shows that as \(A\) increases, mutual information vanishes at progressively smaller separations. Under strong nonlocality (\(A=400\)), it is zero for all separations, indicating the disconnected surface is always the dominant saddle point. This phenomenon is a direct signature of volume-law entanglement, which forces minimal surfaces to remain pinned near the boundary, making the disconnected configuration universally area-minimizing. A comprehensive parameter scan confirms that mutual information is identically zero throughout the entire volume-law regime (\(l \ll A\)), irrespective of separation, and only becomes non-zero when \(l \gtrsim A\).

In summary, holographic mutual information serves as a precise diagnostic: when \(A \gg l\), the vacuum exhibits extensive volume-law entanglement yet displays a complete suppression of bipartite correlations between finite subregions. The parameter \(A\) thus governs a transition between two distinct phases, one with extensive but delocalized entanglement and negligible mutual information, and another with conventional geometric connectivity. This behavior, noted in earlier works \cite{Karczmarek:2013xxa,Pang:2014tpa}, sets the stage for a more profound restructuring of the entanglement architecture.

\subsection{Holographic Tripartite Information and the Decoupled Phase}

\begin{figure}[tbp]
\centering
\includegraphics[width=0.49\textwidth,origin=c]{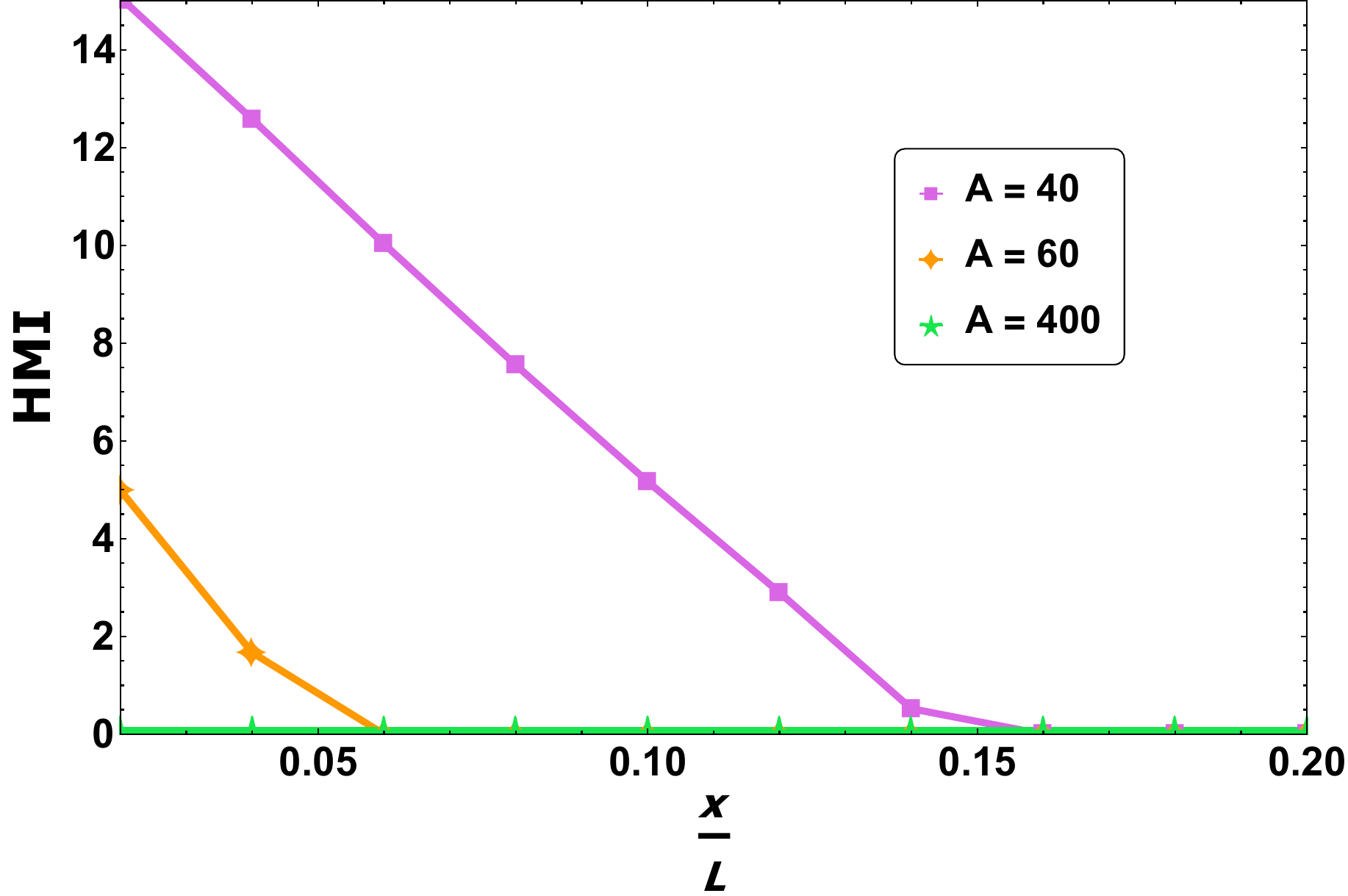}
\includegraphics[width=0.49\textwidth,origin=c]{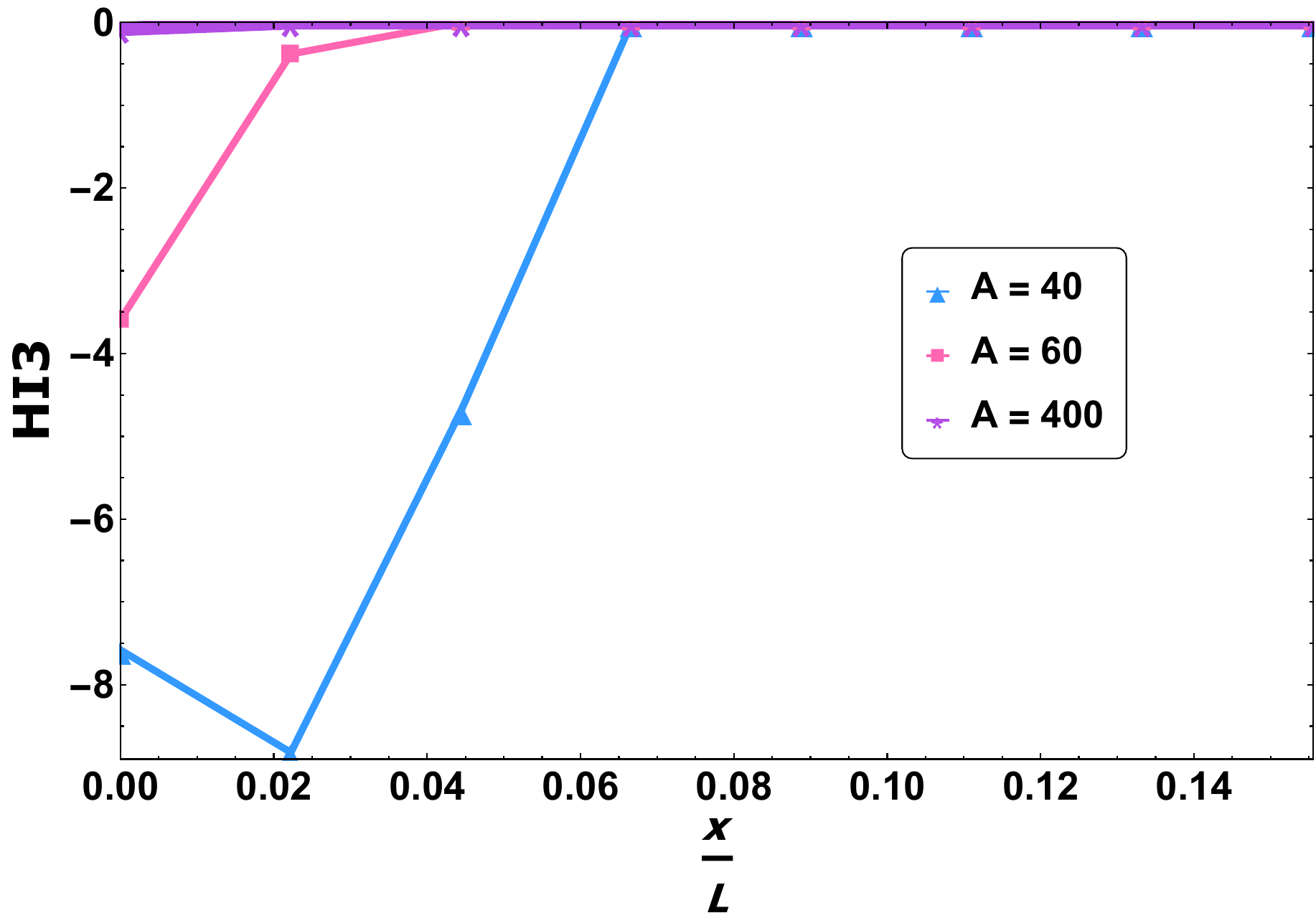}
\caption{\label{fig:7.1} Holographic mutual information (left) and tripartite information (right) for \(d=2\), \(\omega=1\), with equal subsystem lengths \(l=50\), plotted against normalized separation \(x/L\) for \(A=40,60,400\). As \(A\) increases, both mutual information and tripartite information undergo dramatic suppression. For strong nonlocality (\(A=400\)), mutual information and tripartite information almost vanishes for all separations.
 }
\end{figure}

The behavior of the holographic tripartite information, \(I_3\), reveals a systematic departure from established holographic principles, driven by the same geometric mechanism. As shown in Figure \ref{fig:7.1}, \(I_3\) transitions from negative values at moderate nonlocality (\(A=40,60\)), consistent with the monogamy of mutual information \cite{Hayden:2011ag}, to identically zero under strong nonlocality (\(A=400\)). This evolution is starkly at odds with our field-theoretic results, where \(I_3\) remains robustly negative, highlighting a fundamental tension in the strongly nonlocal regime.

The vanishing of \(I_3\) is particularly significant. It represents a saturation of the strong subadditivity bound and signals a controlled departure from the monogamous entanglement structure of conventional holographic systems toward a distinct decoupled phase. In this phase, despite the extensive volume-law entanglement, the specific multipartite correlations quantified by \(I_3\) are entirely absent.

This transition has a clear geometric origin, shared with the suppression of mutual information. The volume-law entanglement, enforced by large \(A\), pins all Ryu-Takayanagi surfaces near the boundary. This prevents the formation of the connected bulk configurations necessary for generating negative \(I_3\) and non-zero mutual information. Consequently, the disconnected surface configuration always dominates for \(A=400\), leading to the simultaneous vanishing of both correlation measures. Thus, the parameter \(A\) functions as a unified geometric switch, tuning the system from a conventional monogamous phase to a novel phase characterized by extensive yet delocalized entanglement that suppresses specific bipartite and multipartite correlations.

%\begin{figure}[tbp]\centering\includegraphics[width=0.8\textwidth,origin=c]{2-MI-A60 a1-15 L1-35,.pdf}\caption{\label{fig:7.2}  Holographic mutual information for \( d=2 \), \( \omega=1 \), and \( A=60 \), shown as a function of both the subsystem length \( l \) and the separation \( x \). For small \( l \) (volume-law regime), the mutual information is zero regardless of \( x \). It becomes non-zero only as \( l \) increases into the crossover region, where it then exhibits the expected decay with increasing separation \( x \).}\end{figure}

\section{Conclusion}

This work extends the foundational results of Shiba and Takayanagi by establishing that the nonlocality scale \(A\) acts as a unified control parameter for both the scaling and the structure of quantum correlations. It drives the transition from area-law to volume-law entanglement entropy while simultaneously reshaping the correlation architecture. The results of numerical and holographic methods reveals a fundamental tension: in the field theory, nonlocality generates persistent long-range mutual information and a distinct form of multipartite entanglement, where increased separation between large subsystems paradoxically deepens their quantum connectivity, as quantified by a more negative \(I_3\). In the holographic dual, however, the same parameter \(A\) triggers a geometric transition that suppresses these very correlations in the volume-law regime. This stark contrast underscores that nonlocality creates complex, non-geometric networks that challenge classical spacetime descriptions.

The holographic dual provides an elegant geometric explanation for the volume-law entropy via a warped bulk spacetime. Though, this success reveals a critical limitation. In the volume-law regime, the model predicts a complete suppression of both mutual and tripartite information, a correlation structure that directly contradicts the rich, monogamous entanglement observed in the field theory. The geometric competition that favors disconnected minimal surfaces fails to capture the intricate web of bipartite and multipartite linkages defining the field theory vacuum.

We found two main results. First, nonlocality is a useful method for creating quantum states with strong, adjustable, long-distance entanglement. Second, to fully describe these systems using holography, we need models that go beyond classical geometry and can capture their full entanglement structure. Although the Ryu-Takayanagi formula remains a foundational element in the study of holographic entanglement, it may be insufficient to fully characterize the intricate entanglement structures present in nonlocal quantum systems. This underscores the necessity for an expanded holographic framework, one that incorporates non-geometric degrees of freedom, to more accurately capture the full spectrum of entanglement phenomena in such contexts.

Future work should focus on extending this analysis to fermionic and gauge-theoretic nonlocal models, probing dynamical and thermal states, and developing refined holographic frameworks capable of capturing the monogamous entanglement and spatial correlation structures that are characteristic of strongly nonlocal field theories.

\subsection*{Note added} We thank Niko Jokela for valuable correspondence, which highlighted the contrast between our findings, specifically, the absence of a phase transition in the holographic entanglement entropy, and the behavior observed in other holographic nonlocal theories \cite{Bea:2017iqt, Jokela:2020wgs}. This discussion helped to clarify the distinctive implications of our geometric setup.

	%%%%%%%%%%%%%%%%%%%%%%%%%%%%%%%%%%%%%%%%%%%%%%%
\subsection*{Acknowledgments}
We gratefully acknowledge Mohsen Alishahiha and M.Javad Vasli for their discussions. Reza Pirmoradian further thanks Ali Mollabashi, M. Reza Mohammadi Mozafar, and Ali Naseh for their valuable early comments and discussions. The work of  Reza Pirmoradian is based on research funded by the Iran National Science Foundation
(INSF) under Project  No. 4026389. Lastly, we recognize the use of AI tools in assisting with text editing and refinement.
%%%%%%%%%%%%%%%%%%%%%%%%%%%%%%%%%%%%%%%%%%%%%%%%


\begin{thebibliography}{99}

\bibitem{Bell:1964kc}   J. S. Bell, On the Einstein Podolsky Rosen paradox, Physics Physique  1, 195 (1964)
doi:10.1103/PhysicsPhysiqueFizika.1.195.




\bibitem{Freedman:1972zza}  S. J. Freedman and J. F. Clauser, Experimental Test of Local Hidden-Variable Theories, Phys. Rev. Lett. 28, 938 (1972)
doi:10.1103/PhysRevLett.28.938.


\bibitem{Hensen:2015ccp} 
B.~Hensen, H.~Bernien, A.~E.~Dreau, A.~Reiserer, N.~Kalb, M.~S.~Blok, J.~Ruitenberg, R.~F.~L.~Vermeulen, R.~N.~Schouten and C.~Abellan, \textit{et al.}
``Loophole-free Bell inequality violation using electron spins separated by 1.3 kilometres,''
Nature \textbf{526}, 682-686 (2015)
doi:10.1038/nature15759
[arXiv:1508.05949 [quant-ph]].


\bibitem{Einstein:1935rr}  A. Einstein, B. Podolsky, and N. Rosen, Can Quantum-Mechanical Description of Physical Reality Be Considered Complete? 
Phys. Rev. \textbf{47}, 777-780 (1935)
doi:10.1103/PhysRev.47.777


\bibitem{Schrödinger:1935} Schrödinger E. Discussion of Probability Relations between Separated Systems. Mathematical Proceedings of the Cambridge Philosophical Society. 1935;31(4):555-563. doi:10.1017/S0305004100013554



\bibitem{Bombelli:1986rw}
L.~Bombelli, R.~K.~Koul, J.~Lee and R.~D.~Sorkin,
``A Quantum Source of Entropy for Black Holes,''
Phys. Rev. D \textbf{34}, 373-383 (1986)
doi:10.1103/PhysRevD.34.373

\bibitem{Srednicki:1993im}
M.~Srednicki,
``Entropy and area,''
Phys. Rev. Lett. \textbf{71}, 666-669 (1993)
doi:10.1103/PhysRevLett.71.666
[arXiv:hep-th/9303048 [hep-th]].


\bibitem{Peschel:2002yqj}
I.~Peschel,
``Calculation of reduced density matrices from correlation functions,''
J. Phys. A \textbf{36}, no.14, L205 (2003)
doi:10.1088/0305-4470/36/14/101
[arXiv:cond-mat/0212631 [cond-mat]].

\bibitem{Eisert:2008ur} Eisert, J., Cramer, M., \& Plenio, M. B. (2010). Colloquium: Area laws for the entanglement entropy. Reviews of Modern Physics, 82(1), 277.
doi:10.1103/RevModPhys.82.277
[arXiv:0808.3773 [quant-ph]].



\bibitem{Wolf:2007tdq}
M.~M.~Wolf, F.~Verstraete, M.~B.~Hastings and J.~I.~Cirac,
``Area Laws in Quantum Systems: Mutual Information and Correlations,''
Phys. Rev. Lett. \textbf{100}, no.7, 070502 (2008)
doi:10.1103/PhysRevLett.100.070502
[arXiv:0704.3906 [quant-ph]].



\bibitem{Groisman:2005dbo}
B.~Groisman, S.~Popescu and A.~Winter,
``Quantum, classical, and total amount of correlations in a quantum state,''
Phys. Rev. A \textbf{72}, no.3, 032317 (2005)
doi:10.1103/PhysRevA.72.032317
[arXiv:quant-ph/0410091 [quant-ph]].

\bibitem{Linden:2002vpn}
N.~Linden, S.~Popescu and W.~K.~Wootters,
``Almost Every Pure State of Three Qubits Is Completely Determined by Its Two-Particle Reduced Density Matrices,''
Phys. Rev. Lett. \textbf{89}, no.20, 207901 (2002)
doi:10.1103/PhysRevLett.89.207901

\bibitem{Hayden:2011ag}
P.~Hayden, M.~Headrick and A.~Maloney,
``Holographic Mutual Information is Monogamous,''
Phys. Rev. D \textbf{87}, no.4, 046003 (2013)
doi:10.1103/PhysRevD.87.046003
[arXiv:1107.2940 [hep-th]].

\bibitem{Koashi:2003pgf}
M.~Koashi and A.~Winter,
``Monogamy of quantum entanglement and other correlations,''
Phys. Rev. A \textbf{69}, no.2, 022309 (2004)
doi:10.1103/PhysRevA.69.022309
[arXiv:quant-ph/0310037 [quant-ph]].

\bibitem{Ryu:2006bv}
S.~Ryu and T.~Takayanagi,
``Holographic derivation of entanglement entropy from AdS/CFT,''
Phys. Rev. Lett. \textbf{96}, 181602 (2006)
doi:10.1103/PhysRevLett.96.181602
[arXiv:hep-th/0603001 [hep-th]].



%\cite{Mirabi:2016elb, Iizuka:2025ioc, Ju:2023tvo, Tanhayi:2017wcd,RezaMohammadiMozaffar:2016lbo,Tanhayi:2016uui}
\bibitem{Mirabi:2016elb}
S.~Mirabi, M.~R.~Tanhayi and R.~Vazirian,
``On the Monogamy of Holographic $n$-partite Information,''
Phys. Rev. D \textbf{93} (2016) no.10, 104049
doi:10.1103/PhysRevD.93.104049
[arXiv:1603.00184 [hep-th]].

%\cite{Iizuka:2025ioc}
\bibitem{Iizuka:2025ioc}
N.~Iizuka and M.~Nishida,
``Genuine multientropy and holography,''
Phys. Rev. D \textbf{112} (2025) no.2, 026011
doi:10.1103/714c-byxq
[arXiv:2502.07995 [hep-th]].
%\cite{Ju:2023tvo}

\bibitem{Ju:2023tvo}
X.~X.~Ju, T.~Z.~Lai, Y.~W.~Sun and Y.~T.~Wang,
``Holographic n-partite information in hyperscaling violating geometry,''
JHEP \textbf{08} (2023), 064
doi:10.1007/JHEP08(2023)064
[arXiv:2304.11430 [hep-th]].

%\cite{Tanhayi:2017wcd}
\bibitem{Tanhayi:2017wcd}
M.~R.~Tanhayi,
``Universal terms of holographic entanglement entropy in theories with hyperscaling violation,''
Phys. Rev. D \textbf{97} (2018) no.10, 106008
doi:10.1103/PhysRevD.97.106008
[arXiv:1711.10526 [hep-th]].

%\cite{RezaMohammadiMozaffar:2016lbo}
\bibitem{RezaMohammadiMozaffar:2016lbo}
M.~Reza Mohammadi Mozaffar, A.~Mollabashi and F.~Omidi,
``Non-local Probes in Holographic Theories with Momentum Relaxation,''
JHEP \textbf{10} (2016), 135
doi:10.1007/JHEP10(2016)135
[arXiv:1608.08781 [hep-th]].

%\cite{Tanhayi:2016uui}
\bibitem{Tanhayi:2016uui}
M.~R.~Tanhayi and R.~Vazirian,
``Higher-curvature Corrections to Holographic Entanglement with Momentum Dissipation,''
Eur. Phys. J. C \textbf{78} (2018) no.2, 162
doi:10.1140/epjc/s10052-018-5620-8
[arXiv:1610.08080 [hep-th]].

\bibitem{Shiba:2013jja}
N.~Shiba and T.~Takayanagi,
``Volume Law for the Entanglement Entropy in Non-local QFTs,''
JHEP \textbf{02}, 033 (2014)
doi:10.1007/JHEP02(2014)033
[arXiv:1311.1643 [hep-th]].




\bibitem{MohammadiMozaffar:2024uiy}
M.~R.~Mohammadi Mozaffar,
``Capacity of entanglement and volume law,''
JHEP \textbf{09} (2024), 068
doi:10.1007/JHEP09(2024)068
[arXiv:2407.16028 [hep-th]].


%\cite{MohammadiMozaffar:2024uiy,MohammadiMozaffar:2017nri,Vasli:2023syq,Khorasani:2023usq}
\bibitem{MohammadiMozaffar:2017nri}
M.~R.~Mohammadi Mozaffar and A.~Mollabashi,
``Entanglement in Lifshitz-type Quantum Field Theories,''
JHEP \textbf{07}, 120 (2017)
doi:10.1007/JHEP07(2017)120
[arXiv:1705.00483 [hep-th]].

\bibitem{Vasli:2023syq}
M.~J.~Vasli, K.~Babaei Velni, M.~R.~Mohammadi Mozaffar, A.~Mollabashi and M.~Alishahiha,
``Krylov complexity in Lifshitz-type scalar field theories,''
Eur. Phys. J. C \textbf{84}, no.3, 235 (2024)
doi:10.1140/epjc/s10052-024-12609-9
[arXiv:2307.08307 [hep-th]].

\bibitem{Doroudiani:2019llj}
M.~Doroudiani, A.~Naseh and R.~Pirmoradian,
``Complexity for Charged Thermofield Double States,''
JHEP \textbf{01}, 120 (2020)
doi:10.1007/JHEP01(2020)120
[arXiv:1910.08806 [hep-th]].	


\bibitem{Khorasani:2023usq}
F.~Khorasani, R.~Pirmoradian and M.~R.~Tanhayi,
``Position dependence of Nielsen complexity for the thermofield double state,''
Phys. Lett. B \textbf{851}, 138585 (2024)
doi:10.1016/j.physletb.2024.138585
[arXiv:2308.15836 [quant-ph]].

\bibitem{Casini:2005zv}
H.~Casini and M.~Huerta,
``Entanglement and alpha entropies for a massive scalar field in two dimensions,''
J. Stat. Mech. \textbf{0512}, P12012 (2005)
doi:10.1088/1742-5468/2005/12/P12012
[arXiv:cond-mat/0511014 [cond-mat.other]].



\bibitem{Azeyanagi:2007bj}
T.~Azeyanagi, T.~Nishioka and T.~Takayanagi,
``Near Extremal Black Hole Entropy as Entanglement Entropy via AdS(2)/CFT(1),''
Phys. Rev. D \textbf{77}, 064005 (2008)
doi:10.1103/PhysRevD.77.064005
[arXiv:0710.2956 [hep-th]].

\bibitem{Herzog:2012bw}
C.~P.~Herzog and M.~Spillane,
``Tracing Through Scalar Entanglement,''
Phys. Rev. D \textbf{87}, no.2, 025012 (2013)
doi:10.1103/PhysRevD.87.025012
[arXiv:1209.6368 [hep-th]].


\bibitem{Herzog:2013py}
C.~P.~Herzog and T.~Nishioka,
``Entanglement Entropy of a Massive Fermion on a Torus,''
JHEP \textbf{03}, 077 (2013)
doi:10.1007/JHEP03(2013)077
[arXiv:1301.0336 [hep-th]].

%\cite{Pirmoradian:2021wvo}
\bibitem{Pirmoradian:2021wvo}
R.~Pirmoradian and M.~R.~Tanhayi,
``Non-local probes of entanglement in the scale-invariant gravity,''
Int. J. Geom. Meth. Mod. Phys. \textbf{18} (2021) no.12, 2150197
doi:10.1142/S0219887821501978
[arXiv:2103.02998 [hep-th]].

%\bibitem{Ghasemi:2021jiy}M.~Ghasemi, A.~Naseh and R.~Pirmoradian,``Odd entanglement entropy and logarithmic negativity for thermofield double states,''JHEP\textbf{10}, 128 (2021)doi:10.1007/JHEP10(2021)128[arXiv:2106.15451 [hep-th]].

\bibitem{Pirmoradian:2023uvt}
R.~Pirmoradian and M.~R.~Tanhayi,
``Symmetry-resolved entanglement entropy for local and non-local QFTs,''
Eur. Phys. J. C \textbf{84}, no.8, 849 (2024)
doi:10.1140/epjc/s10052-024-13212-8
[arXiv:2311.00494 [hep-th]].

\bibitem{Pirmoradian:2025dco}
R.~Pirmoradian and M.~R.~Tanhayi,
``Information Dynamics in Quantum Harmonic Systems: Insights from Toy Models,''
[arXiv:2501.14359 [quant-ph]].


\bibitem{Koffel:2012cu}
T.~Koffel, M.~Lewenstein and L.~Tagliacozzo,
``Entanglement entropy for the long range Ising chain,''
Phys. Rev. Lett. \textbf{109}, 267203 (2012)
doi:10.1103/PhysRevLett.109.267203
[arXiv:1207.3957 [cond-mat.str-el]].	

\bibitem{Nozaki:2012zj}
M.~Nozaki, S.~Ryu and T.~Takayanagi,
``Holographic Geometry of Entanglement Renormalization in Quantum Field Theories,''
JHEP \textbf{10}, 193 (2012)
doi:10.1007/JHEP10(2012)193
[arXiv:1208.3469 [hep-th]].


\bibitem{Karczmarek:2013xxa}
J.~L.~Karczmarek and C.~Rabideau,
``Holographic entanglement entropy in nonlocal theories,''
JHEP \textbf{10}, 078 (2013)
doi:10.1007/JHEP10(2013)078
[arXiv:1307.3517 [hep-th]].

\bibitem{Pang:2014tpa}
D.~W.~Pang,
``Holographic entanglement entropy of nonlocal field theories,''
Phys. Rev. D \textbf{89}, no.12, 126005 (2014)
doi:10.1103/PhysRevD.89.126005
[arXiv:1404.5419 [hep-th]].
	


\bibitem{Bea:2017iqt}
Y.~Bea, N.~Jokela, A.~P{\"o}nni and A.~V.~Ramallo,
``Noncommutative massive unquenched ABJM,''
Int. J. Mod. Phys. A \textbf{33}, no.14n15, 1850078 (2018)
doi:10.1142/S0217751X18500781
[arXiv:1712.03285 [hep-th]].

\bibitem{Jokela:2020wgs}
N.~Jokela and J.~G.~Subils,
``Is entanglement a probe of confinement?,''
JHEP \textbf{02}, 147 (2021)
doi:10.1007/JHEP02(2021)147
[arXiv:2010.09392 [hep-th]].



	
\end{thebibliography}
\end{document}